\documentclass[11pt,twocolumn]{article}
\usepackage{graphicx}
\usepackage[a4paper,margin=1.5cm,top=2cm,bottom=2cm]{geometry}
\usepackage{subfigure}
\usepackage{relsize}
\usepackage{amssymb}
\usepackage{xfrac}
\usepackage{ifthen}
\usepackage{amsmath}
\usepackage{amsfonts}
\usepackage{upgreek}
\usepackage{bm}
\usepackage{multirow}
\usepackage{color}
\usepackage{placeins}

\usepackage{tikz}
\usetikzlibrary{arrows}
\usetikzlibrary{patterns}
\usetikzlibrary{external}
\graphicspath{ {plots/} }

\newcommand{\tot}{\text{inel}}
\newcommand{\inc}{\text{inc}}
\newcommand{\eff}{\text{eff}}
\newcommand{\SPS}{\text{SPS}}
\newcommand{\DPS}{\text{DPS}}
\newcommand{\iSPS}{\text{incSPS}}
\newcommand{\eSPS}{\text{excSPS}}
\newcommand{\iDPS}{\text{incDPS}}
\newcommand{\eDPS}{\text{excDPS}}
\newcommand{\n}{\ensuremath{\bar n}}

\begin{document}

\title{
Double parton scattering with high cross sections
}
\author{
Rafa\l{} Staszewski\thanks{Corresponding author (rafal.staszewski@ifj.edu.pl)}
\\[7mm]
Institute of Nuclear Physics Polish Academy of Sciences\\
ul. Radzikowskiego 152, 31-342 Krak\'ow, Poland\\[7mm]
}
\date{}

\maketitle
\begin{abstract}
The calculations of the double parton scattering cross sections are discussed.
It is shown that the commonly used factorised formula is valid only in the limit of low cross sections. 
The applicability of this approximation is studied with a more general approach based on the Poisson distribution averaged over the impact parameter space.
\end{abstract}

\section{Introduction}
Double parton scattering (DPS) is a relatively new subject of research in hadron physics.
It concerns a class of processes in which two pairs of partons are involved in hard interactions.
This phenomenon is now well established and confirmed by the data at the Tevatron and LHC (see \textit{e.g.} \cite{Abe:1997xk,Abazov:2009gc,Chatrchyan:2013xxa,Aad:2014rua}).

A common way to calculate DPS processes (when both scatterings are of the same type) is the \emph{factorised formula}:
\[
\sigma_\DPS = \frac{1}{2\sigma_\eff} \cdot \sigma_\SPS^2,
\]
where $\sigma_\SPS$ is the cross section of a single parton scattering and $\sigma_\eff$ is a normalisation constant.

It is well known that the factorised formula is approximate, see for example \cite{Diehl:2011yj, Golec-Biernat:2015aza}. It relays on the assumption that the two-parton distribution function, $f^{(2)}(x_1, \mu_1, x_2, \mu_2)$ factorises into the single-parton distributions $f^{(1)}(x, \mu)$:
\[
f^{(2)}(x_1, \mu_1, x_2, \mu_2) = 
f^{(1)}(x_1, \mu_2)\cdot
f^{(1)}(x_2, \mu_2)
\]
This approximation corresponds to a requirement that a presence of one hard scattering does not affect the distributions of the remaining partons.
One can expect it to be valid for partons at small values of $x$, where the densities are large and removing a single parton does not make a large difference.

Neglecting possible correlations between the two partons in a hadron is not the only drawback of the factorised formula.
As will be argued in the present paper, the formula is applicable only to processes with small cross sections.
A hint towards this conclusion can already be seen in \cite{Luszczak:2011zp}, where DPS charm production at the LHC is studied and 
it is shown that the factorised formula can predict DPS cross sections that exceed the total inelastic cross section.
This suggests some limitations of the theoretical framework used in the calculations, which will be explored in the present study.

The paper is organised as follows. In Section \ref{sec:toy} a simplistic model of the problem is presented, in which it it possible to discuss the key ideas.
Section \ref{sec:real} presents a more realistic approach, which can be used is actual calculations.
The numerical results are presented in Section \ref{sec:num} and Section \ref{sec:sum} contains a summary.

\section{Toy model}
\label{sec:toy}

The first step for a correct description of the DPS processes is the proper interpretation of the inclusive cross section.
It is a standard quantity calculated as a convolution of parton densities with a parton-level cross section
\[
\sigma_\inc = \int f(x_1, \mu) f(x_2, \mu) \hat\sigma(x_1 x_2, \mu)\,\text{d}x_1 \, \text{d}x_2.
\]
$\sigma_\inc$ should be understood as a cross section for the given \emph{subprocess}, \textit{i.e.} a \textit{parton--parton} interaction.
It is important to distinguish this from a cross section for a \emph{process}, \textit{i.e.} a \textit{hadron--hadron} interaction.

For a given \textit{subprocess} it is possible to define various processes based on the number of subprocesses they contain:
\begin{itemize}
  \item \textbf{inclusive SPS}: \emph{at least one} subprocess,
  \item \textbf{exclusive SPS}: \emph{exactly one} subprocess,
  \item \textbf{inclusive DPS}: \emph{at least two} subprocesses,
  \item \textbf{exclusive DPS}: \emph{exactly two} subprocesses,
  \item \textit{etc}
\end{itemize}
(a similar naming convention can also be found in~\textit{e.g.}~\cite{Rogers:2008ua}).

Very often the distinction between a \textit{process} and a \textit{subprocess} is superfluous, but for (sub)processes with high cross sections it in fact is essential.
In particular, the cross section for a \textit{subprocess} ($\sigma_\inc$) can exceed the total inelastic cross section.
This corresponds to the case when on average more than a single subprocess takes place in the hadron--hadron process.
On the other hand, the cross section for an (inelastic) \textit{process} must, by definition, always be smaller than the total inelastic cross section.
These ideas have been already noticed in \cite{Sjostrand:1987su} and are the basis for the modeling of the underlying event in the contemporary Monte Carlo event generators.

A useful quantity for the following derivations is the average number of subprocesses in a process: \n.
It is related to the inclusive cross section and the total inelastic cross section by:
\[
\sigma_\inc = \n \sigma_\tot.
\]
In the present simplistic toy model it is assumed that the actual number of subprocesses, $n$, follows a Poisson distribution with average $\n$.
The probability of a given number of subprocesses is given by:
\[
P\left(n\right) = e^{-\n} \frac{ \n^n }{n!}
\]

With the formula for the probability, it is straightforward to calculate cross sections for different processes:
\begin{itemize}
  \item \textbf{inclusive SPS}:
    \[
    \sigma_\iSPS = P(n\ge1) \cdot \sigma_\tot = [1-P(0)] \cdot \sigma_\tot,
    \]
  \item \textbf{exclusive SPS}:
    \[
    \sigma_\eSPS = P(1) \cdot \sigma_\tot,
    \]
  \item \textbf{inclusive DPS}:
    \[
    \sigma_\eDPS 
    = P(n\ge 2) \cdot \sigma_\tot 
    = [1 - P(0) - P(1)] \cdot \sigma_\tot,
    \]
  \item \textbf{exclusive DPS}:
    \[
    \sigma_\iDPS = P(2) \cdot \sigma_\tot.
    \]
\end{itemize}

It is worth deriving the small cross section (small \n) limit of the above formulae.
For $\n\to0$,  $\sigma_\iSPS$ and $\sigma_\eSPS$ both tend to $\sigma_\inc$. This explains why in typical situations distinguishing between these three, in principle different, quantities is not necessary.
In the same limit, the inclusive and exclusive DPS cross sections can both be approximated by
\[
\frac{1}{2\sigma_\tot} \cdot \sigma_\inc^2,
\]
which has a form of the factorised formula with $\sigma_\eff = \sigma_\tot$.

\section{Realistic model}
\label{sec:real}

The realistic model of multiple parton scattering is based on the same principles as  above,
with an additional complication originating from the geometry of the collisions.
It is assumed that the number of interactions follows a Poisson distribution, like in the toy model, but with the average value depending on the impact parameter of the interaction,~$b$:
\[
\n \to \n(b),
\]
\[
P(n) \to P(n; b) = e^{-\n(b)} \frac{[\n(b)]^n}{n!}.
\]

Assuming the knowledge of $\n(b)$, it is straightforward to calculate cross section for given processes, for example:
\[
\sigma_\iSPS = \int [1 - P(0; b)] \,  \text{d}^2\boldsymbol{b}.
\]
In turn, the calculation of inclusive cross section should take into account all the occurrences of a given subprocess, therefore
\[
\sigma_\inc = \int n(b) \, \text{d}^2 \boldsymbol{b}.
\]
This equation suggests the introduction of the following factorisation (which is always possible):
\[
n(b) = \sigma_\inc F(b)
\]
where $F(b)$ is known as the overlap function.
It can be interpreted as the transverse distribution of interactions, and it is normalised to unity:
\[
\int F(b) \, \text{d}^2 \boldsymbol{b} = 1.
\]

It is worth studying the limit of small $\n(b)$ (or small $\sigma_\inc$) of the above formulae.
It is easy to check that, similarly to the toy model, the $\sigma_\inc$, $\sigma_\iSPS$ and $\sigma_\eSPS$ become equal to each other.
The formulae for $\sigma_\iDPS$ and $\sigma_\eDPS$ both become equal to:
\[
\sigma_\text{DPS} = \frac{1}{2} \sigma_\inc^2 \int F^2(b)\, \text{d}^2\boldsymbol{b}.
\]
Assuming that $F(b)$ has the same form for all processes, the above equation is exactly the standard factorised formula with a universal effective cross section given by:
\[
\frac{1}{\sigma_\eff} = \int F^2(b)\, \text{d}^2\boldsymbol{b}.
\]
This formula provides the probabilistic interpretation of the effective cross section: it is the inverse of the average (the expectation value of) the overlap function.

\section{Numerical results}
\label{sec:num}

In order to perform numerical calculations one needs to assume some shape of the overlap function.
A natural choice is that of a Gaussian with a unit normalisation and a width expressed by the effective cross section (which value is experimentally known to be close to 15~mb at the Tevatron and LHC energies\footnote{The present analysis shows that the factorised formula is not sufficient to describe double parton scattering processes. Since it has been used to extract the effective cross section from experimental data, the obtained values may not be fully correct. However, since this extraction was based on processes with relatively small cross section, one can safely take the present $\sigma_\eff$ value to estimate the magnitude the discussed effects.}):
\[
F_\text{Gaus}(b) = \frac{2}{\sigma_\eff} \exp\left( - \frac{2\pi b^2}{\sigma_\eff} \right).
\]

\begin{figure}[htbp]
  \centering
  \includegraphics[width=\linewidth,page=7]{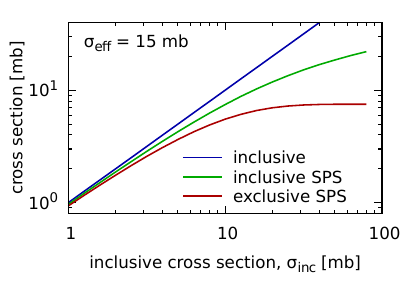}
  \caption{Cross sections for inclusive and exclusive SPS as a function the inclusive cross section.}
  \label{fig:real_sps}
  \centering
  \includegraphics[width=\linewidth,page=8]{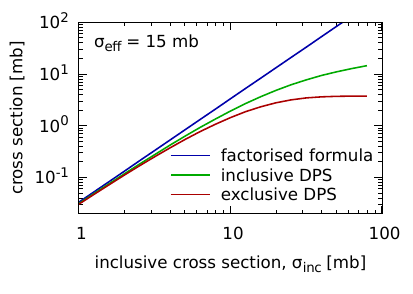}
  \caption{Cross sections for factorised formula, inclusive and exclusive DPS as a function of $\sigma_\inc$.}
  \label{fig:real_dps}
  \centering
  \includegraphics[width=\linewidth,page=9]{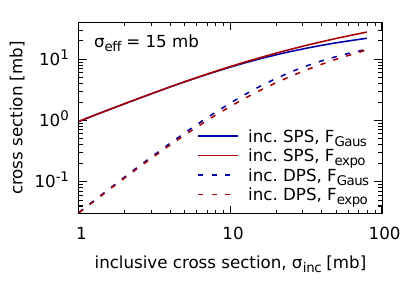}
  \caption{Comparison of cross sections for single and double parton scattering obtained with different shapes of the overlap function.}
  \label{fig:real_shape}
\end{figure}

Figure \ref{fig:real_sps} presents the results obtained for the single parton scattering processes.
In the regime of small cross sections both the exclusive and inclusive SPS cross sections are close to $\sigma_\inc$.
However, once the cross sections reach the order few millibarns, the differences start to become significant and become very large for $\sigma_\inc > 10\ \text{mb}$.

The situation for the double parton scattering is presented in Fig. \ref{fig:real_dps}.
For $\sigma_\inc < 1$~mb, the cross sections for inclusive and exclusive DPS are very close to the cross section obtained with the factorised formula.
For higher cross sections the differences increase and can exceed one order of magnitude for $\sigma_\inc \sim 100\ \text{mb}$.

Finally, it is worth studying how the above results depend on the assumed shape of the overlap function.
A possible choice of an alternative shape is the exponential function (the parameters were chosen to reproduce the normalisation and the effective cross section):
\[
F_\text{expo}(b) = \frac{4}{\sigma_\eff} \exp\left(-b\sqrt\frac{8\pi}{\sigma_\eff}\right).
\]
The comparison of inclusive SPS and DPS cross sections calculated with two different overlap functions is presented in Fig. \ref{fig:real_shape}.
Despite a significantly different shape of the assumed $F(b)$, the predicted cross sections are very similar.

\section{Summary}
\label{sec:sum}

It has been argued that the factorised formula for calculations of double parton scattering processes is valid only in the limit of small cross sections. In a general case it is crucial to properly introduce statistical effects.
The discussed approach is based on the observation that the inclusive cross section (convolution of the partonic cross section with appropriate parton densities) can be related to the average number of parton--parton interactions of a given kind in a hadron--hadron collision.

A set of formulae that can be used to calculate cross sections for processes with different multiplicities of parton scatterings was given.
The discussed approach is, for processes with small cross sections, in agreement with the commonly used factorised formula.
However, once the cross sections become comparable with the total inelastic cross section, these two approaches give significantly different results.
The factorised formula can overestimate the cross section even by an order of magnitude.
It has been shown that the predictions depend very mildly on the chosen shape of the overlap function, as long as its expectation value (\textit{i.e.} the effective cross section) is properly reproduced.

\section*{Acknowledgments}

This work was supported in part by Polish National Science Center grant no. 2015/19/B/ST2/00989.

\end{document}